\documentclass[twocolumn,showpacs,preprintnumbers,amsmath,amssymb]{revtex4}
\usepackage{graphicx}
\usepackage{dcolumn}
\usepackage{bm}

\begin{document}

\preprint{APS/123-QED}

\title{  G\"odel's universe  and the chronology protection conjecture }
\author{P.Pitanga}
\affiliation{Universidade Federal do Rio de Janeiro. Instituto de
F\'{\i}sica. Caixa Postal 68528 \\
Cidade Universit\'aria. 21945-970. Rio de Janeiro, Brazil }
\email{pitanga@if.ufrj.br}

\date{\today}

\begin{abstract}
 We present a  solution for the geodesic motion in  G\"odel's universe  that provides a par\-ti\-cu\-lar proof of Hawking's chronology protection conjecture in three-dimensional gravity theory. The solution is based upon the fact that the seven-dimensional group of the automorphisms of the Heisenberg motion group $H^1 \times U(1)$, modulo discrete sub-group ${\bf Z}$, act isometrically on the boundary of the hyperbolic three-dimensional manifold.Closed timelike curves do not exist due to the  presence of a closed  Cauchy-Riemann surface for chronology protection, with two mirror symmetric sets of helicoidal self-similar modules inside.The present solution is isometrically equivalent to a cylindrical gravitational monochromatic  wave front.
\end{abstract}
\maketitle 

 In 1949, G\"odel \cite {KU} discovered a new solution for Einstein's field equations in which causality is violated by the existence of  periodic world lines that run back into themselves. The periods are given by  integrals over the proper time differential $ds$. These cycles are known as closed timelike curves (CTCs). According to Tipler \cite {TI} a massive infinite  rotating cylinder should create a frame dragging effect in the space-time giving rise to CTCs. Tipler suggested, without proof, that in a finite rotating cylinder CTCs would arise allowing travel into the past.

The non existence of CTCs in acceptable spacetimes, was demonstrated by Deser, Jackiw and 't Hooft \cite {DD} in 1992. In the same year,Hawking \cite {HH}  showed that, according to the ge\-ne\-ral theory of relativity, it is impossible to build a time machine in finite regions of the spacetime without curvature singularities, due to the presence of negative energy states.Hawking posited the existence of a Cauchy-Riemann spacelike surface with a compact horizon for chronology protection, so that causality violation could never be observed from the outside. This conjecture is known as the chronology protection conjecture. For the chronology protection conjecture see also \cite {SUR,MI} 

Nowadays it is believed that some  fundamental pro\-per\-ties of supermassive black holes may be understood  in the realm
 of a $(2n+1)$-dimensional G\"odel Universe. In particular, the static Gimon-Hashimoto solution for the $n=2$ case \cite {HA}, has attracted considerable attention associated with  $M$-theory, pp waves, thermodynamics of black holes and strong gravitational lensing effects and has been the focus of recent research \cite {SQ,SC,CU}.  However, the geodesic motion in G\"odel's universe is not yet completely understood even though great strides have been made over the past $55$ years \cite{KA,CH,IV,TE,IS,DA}.

The aim of this letter is to  provide, the appropriate geometrical setting to prove Hawking's chronology protection conjecture in  G\"odel's Universe. For our purpose it is sufficient to analyze the $(2+1)$-dimensional G\"odel universe in a finite rotating cylinder with constant angular velocity ${\omega}$, and negative cosmological constant ${\Lambda}$.

 First of all, we show that G\"odel's spacetime  is invariant under local actions of the $7$-dimensional group of  automorphisms of the reduced Heisenberg motion group $G=H^1/{\bf Z}\times U(1)$. Thus, the geodesic problem  is reduced to a sub-Riemannian variational problem in the cotangent  space, $T^{*}{\cal M}$, of the $3$-dimensional Heisenberg group, a well known problem in the realm of ma\-the\-mati\-cal control theory \cite {JU,P3}.

{\it The Heisenberg Group} - The Heisenberg group  plays the fundamental role in diverse topics such as har\-mo\-nic analysis \cite {OW}, classical and quantum mechanics   \cite {GB}, and sub-Riemannian geometry (also known as Carnot-Charath\'eodory geometry) \cite {MO,PP,P2,GR}. As it is the most simple non-Abelian nilpotent Lie group homeomorphic to the Euclidean space it offers the  opportunity  for ge\-ne\-ra\-li\-zing the remarkable results of non-commutative harmonic analysis to soluble models of non-Abelian gauge field theories. In particular, it offers a interesting soluble scenario for  Gribov's theory of quark confinement\cite {GRI}

In 2000,the Heisenberg group was used  to prove the possible existence of  bound pairs of self-dual non-Abelian magnetic monopoles  in high energy physics \cite {P1}. Recently, experiments with  superconductors (SCs) doped with topological insulators (TIs), confirms the exis\-tence of such unusual configuration \cite {RT}.

Recall that the  $
(2n+1)$-dimensional Heisenberg group $H^{n}$, can be realised as the multiplicative group of real
  $(n+2)\times (n+2)$ matrices of the form 
   
\begin{equation}A(x,y,t)=\left( \begin{array}{ccc}
1& x&t\\
0&1& y\\
0&0&1
\end{array}\right)\equiv  (z,t)\in C^n_{\times}\times {\Re},
\end{equation}
 
\noindent were we have identified ${\Re}_{\times}^n\times {\Re}_{\times}^n$ with $C_{\times}^n$. By setting $z=x+iy$ and $w=u+iv$, $H^n$ can be defined by the group law
\begin{equation}
\label{UY}
(z,t)\cdot (w,s)=(z+w,t+s +\frac{1}{2}Im(z\cdot {\overline w})),
\end{equation}
where $Im(z\cdot {\overline w})=(u\cdot y -v\cdot x)$ is the standard symplectic form on ${\Re}^{2n}$ and $z\cdot {\overline w}=z_1{\overline w}_1+\ldots + z_n{\overline w}_n$ is the standard Hermitian form on $C^n$.

The $(2n+1)$ left invariant vector fields 
\begin{equation}
Q_j=\frac {{\partial}}{{\partial}_{x_j}}-\frac {1}{2}y_j\frac {{\partial}}{{\partial}_{t}}; P_j=\frac {{\partial}}{{\partial}_{y_j}}+\frac{1}{2}x_j\frac {{\partial}}{{\partial}_{t}};\; T=\frac {{\partial}}{{\partial}_{t}}
\end{equation}

\noindent are the generators of the Heisenberg Lie algebra $h^n$:
\begin{equation}
\label{PE}
[P_i,P_j]=[Q_i,Q_j]=0;[Q_i,P_j]={\delta}_{ij}T.
\end{equation}
 The Heisenberg motion group $ G=H^n\times U(n)$, is the group of isometries of the sub-Laplacian ${\cal L}$: 

\begin{equation}
\label{BU}
{\cal L}=-\sum_{j=1}^{n}(P^2_j+Q^2_
j)=-{\Delta}_z-\frac{1}{4}{\bf r}^2{\partial}_t^2+J{\partial}_t,
\end{equation}
\noindent where ${\Delta}_z$ is the Laplacian on $C_{\times}^n$, ${\bf r}^2=|z|^2$  and 
\begin{equation}
J= \sum _{j=1}^{n}(x_j{\partial}_{y_{j}}-y_j{\partial}_{x_{j}})
\end{equation}
is the rotation operator. In order to have a  bounded  space-time we must work with  the compact group  $H^1/{\bf Z}$. The  discrete twisted  sub-group ${\bf Z}$ is generated by the set of three matrices ${\bf Z}=\{(1,0,0),(0,1,0),(0,0,1/2k{\pi});k=\pm 1,\pm 2,\pm 3,...\}$ \cite {GE}. 

{\it G\"odel's Universe }
  is a rotating anisotropic  homogeneous Lorentzian spacetime, in which  matter takes the form of a pressure-free perfect fluid ($T_{ab}={\rho}{\bf u}_a{\bf u}_b)$ where ${\rho}$ is the matter density and ${\bf u}_a$ are the four normalized velocity vector fields. The  manifold is ${\cal M}=\Re^3\times \Re$, with metric 
\begin{equation}
\label{S}
ds^2=-c^2dt^2 +dx^2 -\frac{1}{2}e^{2{\sqrt 2}kx}dy^2-2e^{{\sqrt 2}kx}cdtdy +du^2 
\end{equation}
where 
 $ k={\omega}/{c}$. The relevant manifold is ${\Re}^3$.  
 
Einstein's field equations are satisfied if $|{\bf u}_{0}|=|{\bf u}_{4}|=1$ and $4{\pi}{\rho}=k^2=-{\Lambda}$. 
 The constant $k$ characterizes the angular velocity ${\omega}$ of the matter associated to the flow vector field ${\bf u}_{4}$, Cf \cite {HW}.

It is easy to verify (taking into account $({\omega},c)$), that (\ref {S})  is invariant under the  automorphisms of the Heisenberg group. For $Aut(H^n)$ see \cite {GB} 

Indeed for each
  ${\alpha}_i(x,y,t,u)\equiv{\alpha}_i ({\bf x})$,${\alpha}_i\in Aut(G)$, we have :

\vspace{0.15cm}
\noindent 1) Symplectic actions 
${\alpha}_1({\bf x})=(y,-x,t,u)$; 

\noindent 2) Dilations 
 ${\alpha}_2({\bf x})=(ax,ay,a^2t,au)$;$k\rightarrow k/a;a>0$;

\noindent 3)Temporal Inversions 
${\alpha}_3({\bf x})=(x,y,-t,u)$; 

\noindent 4) Spatial Inversions
${\alpha}_4({\bf x})=(-x,-y,t,-u)$;

\noindent 5) ${\bf Z}/2{\bf Z}\oplus{\bf Z}/2{\bf Z}$ actions
 ${\alpha}_5({\bf x})=(-x,-y,t,u)= (-x,y,-t,u)= (x,-y,-t,u)$;

\noindent 6)Rotations
 ${\alpha}_6({\bf x})=({\sigma }(x,y),t,u);\; {\sigma}\in U(1)$ 

\noindent 7)Screw translations:
${\alpha}_7({\bf x})=({\bf R}_{\theta}({\bf r}),t+f({\theta}),u)$,
where ${\bf R}_{\theta}\in SO(2);\;{\bf r}^2=x^2+y^2$ and $f({\theta})$ an harmonic function.

 This invariance is the result of the fundamental  principle connecting  the Cauchy-Riemann geometry of the  Poincar\'e  unity ball $B_{n+1}$ ( the Siegel half-space $S_{n+1}$) and the sub-Riemannian geometry, which arises from the correspondence between the boundary  of the Poincar\'e  ball  and the Heisenberg group \cite {JO}.
So this principle enables us to identify the reduced Heisenberg motion group with the boundary of the finite $(2+1)$ - dimensional G\"odel Universe.

 The most symmetric metric of the   Heisenberg motion group is
\begin{equation}
\label{A}
ds^2= dx^2 +dy^2 +du[dt +g(xdy-ydx)] +du^2
\end{equation}
\noindent 

\noindent where $g=({\sqrt 2}/2){\omega}$, in unities of $c=1$.

The contact form  ${\Omega}=dt +g(xdy-ydx)$ is  the annihilator of the vector fields $(Q,P)$, spans of the horizontal sub-space of the tangent space $T{\cal M}=H\oplus V$:
\begin{equation}
 Q={\partial}_x-gy{\partial}_t; P={\partial}_y+gx{\partial}_t,
\end{equation}
\noindent and the dual of the vector fields, $(T,J)$, spans of the vertical sub-space :
\begin{equation}
T={\partial}_t ;\;\;J=g(x{\partial}_y-y{\partial}_x).
\end{equation}
\noindent That is,
\begin{equation}
Q({\Omega})=P({\Omega})=0;
T({\Omega})=1; J({\Omega})=g^2(x^2+y^2).
\end{equation}
Einstein's field equations are satisfied if $|T|=|J|=1$. Hence,
\begin{equation}
\label{OI}
g^2(x^2+y^2)=1\rightarrow  {\omega}= \frac{{\sqrt 2}}{{\bf r}_0};\rightarrow {\Lambda}=-\frac {2}{{\bf r}^2_0}. 
\end{equation}

The  integral lines of the vector fields $(Q,P,T,J)$  are solutions of Einstein field equations. The solutions  depend on G\"odel's radius, ${\bf r}_0$, which characterizes the mass density and the angular velocity. The  circles $C$ at $t=constant$ are called timelike circles if ${\bf r}>{\bf r}_0$,  nullcircles , if ${\bf r}={\bf r}_0$  and spacelike circles,  if ${\bf r}<{\bf r}_0$, Cf \cite {IS}. 

In $T^{*}{\cal M}$, the 
 vector fields $(Q,P,T,J)$ can be re\-pre\-sen\-ted  by the functions: 
\begin{equation}
Q=p_x-gyp_t;P=p_y+gxp_t ;T=p_t;J=g(xp_y-yp_x)
\end{equation}

An explicit calculation gives us the Lie algebra:
\begin{equation}
\{Q,P\}=2gT;\{J,P\}=gQ;\{J,Q\}=-gP 
\end{equation}

\begin{equation}
\{P,T\}=\{Q,T\}=\{J,T\}=0.
\end{equation}

This  Lie algebra  arises in the Nappi-Witten mo\-del of a four-dimensional homogenous anisotropic spacetime, based on nonsemisimple Lie groups, where the me\-tric (\ref {A}) was interpreted as the metric of a gravitational monochromatic plane wave \cite{NA}.

 It will be shown in the following that the metric (\ref {A}) gives us the wave front of a  cylindrical gra\-vi\-ta\-ti\-onal wave, first considered  by Einstein \& Rosen  \cite {AE} and latter des\-cri\-bed in details by Weber \& Wheeler \cite {WE}, and  Marder \cite {MAR}.

The  geodesics are obtained here from the Hamiltonian
\begin{equation}
 {\cal H}=\frac{1}{2}(Q^2+P^2+T^2)+J^2
\end{equation}
 
The Hamiltonian equations are
\begin{mathletters}
\begin{equation}
\label{L}
\frac {dQ}{ds}=\{{\cal H},Q\}=P\{P,Q\}+2J\{J,Q\}=-2g{\lambda}P,
\end{equation}
\begin{equation}
\frac {dP}{ds}=\{{\cal H},P\}=Q\{Q,P\}+2J\{J,P\}=2g{\lambda}Q,
\end{equation}
\begin{equation}
\frac{dT}{ds}=\{{\cal H},T\}=0;\;\;\;
\frac{dJ}{ds}=\{{\cal H},J\}=0,
\end{equation}
 with the supplementary conditions
\begin{equation}
\label{P}
\frac {dx}{ds}=Q;\;\;\;
\frac {dy}{ds}=P;\;\;\;
\frac {dt}{ds} +g\left(y\frac {dx}{ds}-x\frac {dy}{ds}\right)=0,
\end{equation}
\end{mathletters}

Introducing 
$Q={\bf r}_0cos{\psi},P={\bf r}_0sin{\psi}$,
we have 
\begin{equation}
-2g{\lambda}P=\frac {dQ}{ds}=-{\bf r}_0sin{\psi}\frac {d{\psi}}{ds}.
\end{equation}
It follows that ${\psi}= 2g{\lambda}s+{\phi}$, where ${\phi}$ gives the initial directions of the sub-Riemannian geodesics emanate from the origin $(0,0,0)$ of the spacetime.
 Integrating (\ref {P}), in the interval $(0,1)$, one obtains
 the sub-Riemannian wave front, that is, the manifold of the endpoints of the sub-Riemannian geodesics of length ${\bf r}_0$ .

\begin{mathletters}
\begin{eqnarray}
\label{u}
x={\bf r}_0\left[\frac{cos({\theta} +{\phi})-cos{\phi}}{{\theta}}\right],\\
y={\bf r}_0\left[\frac{sin({\theta} +{\phi})-sin{\phi}}{{\theta}}\right],\\
 t={\bf r}_{0}^2\left[\frac{{\theta} -sin {{\theta}}}{{\theta}^2}\right].
\end{eqnarray}
\end{mathletters}
\noindent where ${\theta}= 2g{\lambda}$, with ${\lambda}\in {\Re}$ and  $0\leq {\phi}\leq 2{\pi}$. 
These parametric equations, aided by computers, give us
  all relevant information about the geometry  and topology of the $(2+1)$-dimensional G\"odel's spacetime manifold:
\begin{enumerate}
\item The C circles are given by ${\bf r}=({\bf r}_0/{\theta})\sqrt {2(1-cos{\theta})}$
\item The horizons are localized at $\pm {\pi}\leq {\theta}\leq \pm 2{\pi}$, with the same radius ${\bf r}_h=2{\bf r}_o/{\pi}$.
\item The conic singularities are {\it localized} at ${\theta}= \pm 2k{\pi}$  in the points $({\bf r}=0,t_k={\bf r}_0^2/2k{\pi});\;\;\;k=\pm 1,\pm 2,...$.
\item There are no  timelike circles. The maximal circle is the nullcircle of radius ${\bf r}={\bf r}_0$ in the plane $t=0$ of the Cauchy-Riemann surface at ${\theta}=0$
\item The boundary is a incomplete manifold, axially symmetric, formed by a infinite set of rectifiable curves, as shown in fig.1.
\begin{figure}
\begin{center}
\includegraphics[width=0.2\textwidth]{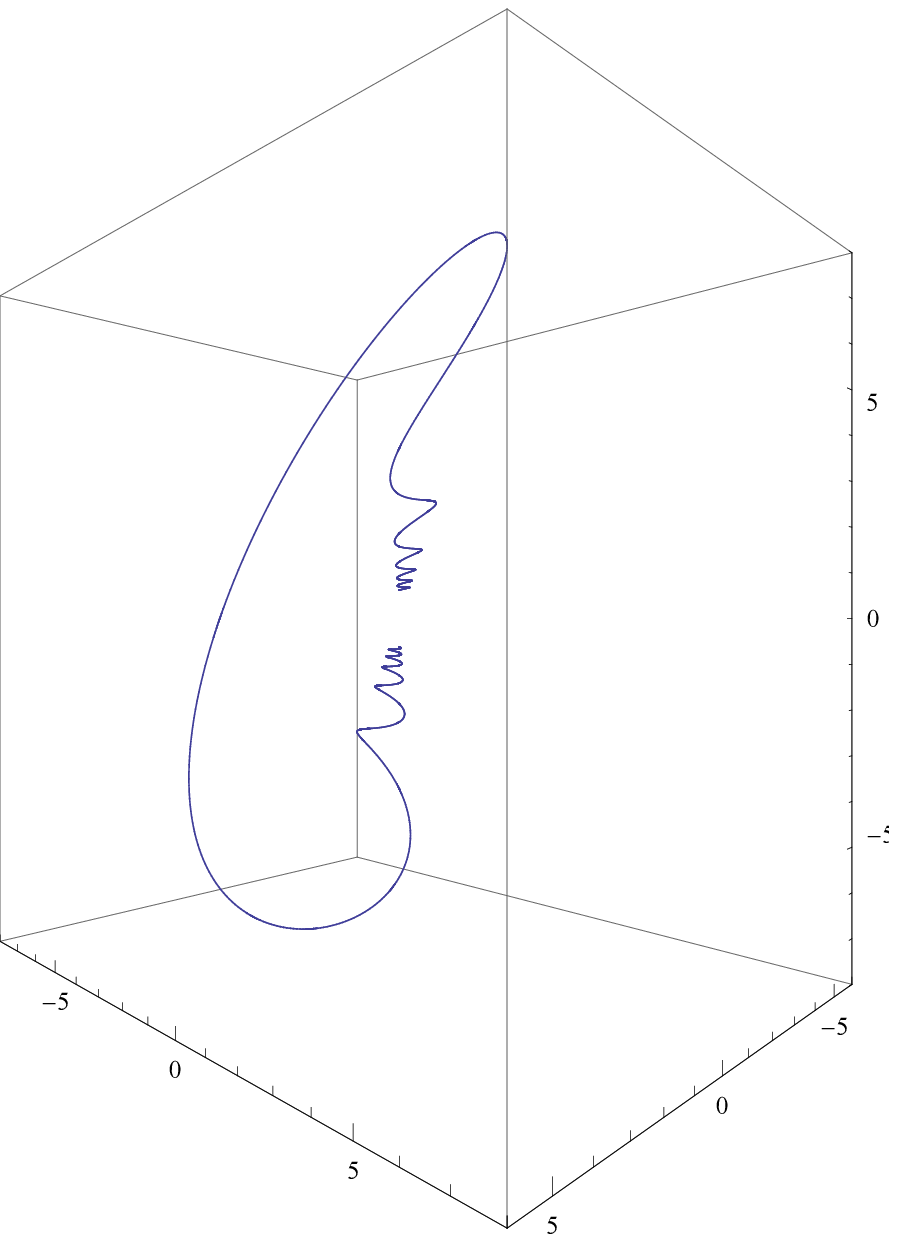}
\caption{\label{fig:wide1}} 
\end{center}
\end{figure}
\item The Cauchy-Riemann surface  with the array of self-similar modules is shown in fig.2.
\begin{figure}
\begin{center}
\includegraphics[width=0.2\textwidth]{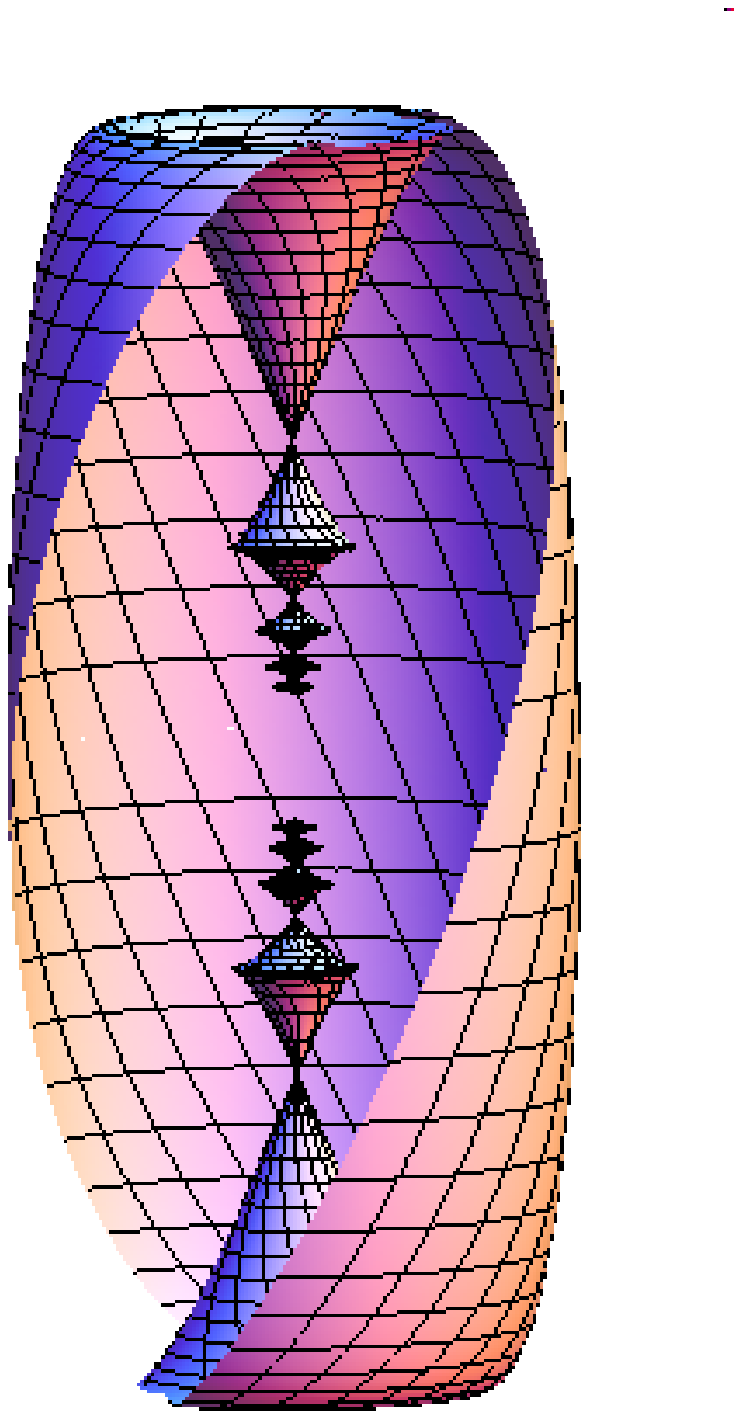}
\caption{\label{fig:wide1}} 
\end{center}
\end{figure}
\item The first module localized at $2{\pi}\leq {\theta} \leq 4{\pi}$ is shown in fig.3.
\begin{figure}[!h]
\begin{center}
\includegraphics[width=0.2\textwidth]{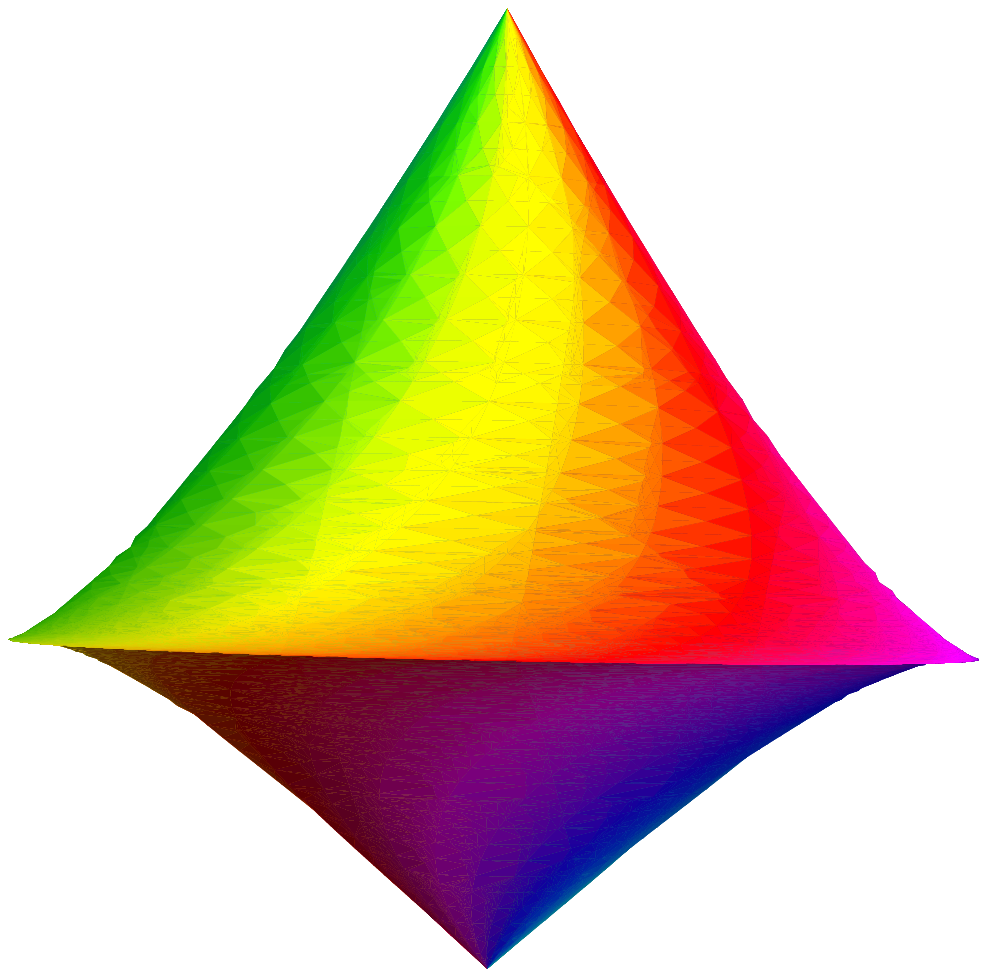}
\caption{\label{fig:wide1}} 
\end{center}
\end{figure}

\end{enumerate}

{\it Conclusions} - In this   letter we have presented a finite cylindrical  solution for the geodesic motion in the $(2+1)$ - dimensional G\"odel universe, based on the geometry and topology of the compact Heisenberg motion group. Causality is not violated, due to the existence of a well  defined Cauchy-Riemann spacelike surface with compact  dynamically-generated conic horizons.The spacetime is causally geodesically incomplete (according to Hawking-Penrose conditions \cite {JS}) and strongly supports Hawking's chronology protection conjecture:{\it the laws of physics do not allow the appearance of closed timelike curves}.

There is no room for negative energy (mass). As it is well know, in $3$-dimensional Einstein gravity, the only possible energy  measure must be topological and the Euler invariant is the only candidate. From the Gauss-Bonnet theorem we have $K{\sum}_{n=0}^{\infty}\int_{A_n}d{\Omega}=4{\pi}e({\Sigma})$, where $K$ is a constant, $ A_n$ is the area enclosed by each maximal  circle, $C_n$, in each region of the spacetime, and $e({\Sigma})$, is the Euler number of the $2$-dimensional {\it orbifold}, ${\Sigma}={\Sigma}^{-}\cup {\Sigma}^{+}$. As is well known, after Thurston \cite {TU}, in the case of a closed $3$-manifold with  geometric structure modelled on the compact Heisenberg group $H^1/{\bf Z}$, (a manifold foliated by circles), $e({\Sigma})=1$. 

The mass, at the origin of the spacetime, only af\-fe\-cts geometry globally rather than locally.The global geometry is fixed by periodic conic singularities along the source's world line, according to the three-dimensional gravity theory of Deser, Jackiw and 't Hooft \cite {DE}. 
This closed homogeneous anisotropic spacetime shares many interesting properties with a physically acceptable spacetime. However, due to the notable absence of the Riemann tensor, (replaced by the non-holonomic connection, ${\Omega}$), G\"odel's universe based on the sub-Riemannian geometry of the Heisenberg group, is not ac\-ce\-pta\-ble at the large scale structure of spacetime of Einstein's ge\-ne\-ral theory of relativity. Note that the rotating body (fig.2), cannot be embedded in ${\Re}^3$, as a surface of revolution as is usual in  Riemannian geometry. Indeed, the set of end points of the sub-Riemannian geodesics emanate from the origin, is a convex set in ${\Re}^3$ whose boundary has two opposed enumerable sets of conic singularities. The only really singular point is the unreachable origin, which is the closure of the spacetime.

 However, this unusual configuration is acceptable at the small scale structure of quantum theory, according to the fundamental Stone-Von Neumann theorem on the unitary and irreducible representation of the Heisenberg group in the Hilbert space, which is fairly simple and well understood \cite {OW,GB}.

Finally it should be noted that there is a close connection between this approach and \cite {DD}, where the flat Kerr metric was used to represent the spacetime.Indeed, the three-dimensional manifold of the compact Heisenberg group, $H^1/{\bf Z}$, is a line  bundle over a plane,in ${\Re}^3$, with a prescription for identifying points in ${\Re}^3$. This $3$-dimensional topology was used to give an example of how to close a Cauchy-Riemann surface:

{\it "The matching condition is defined only when a closed curve is followed around the source ;these matchings are defined  by a deficit angle and a time shift. More precisely we have a space with ${\Re}^3/{\Re}$ topology. (a three-space  with a line obstruction on the source's world line) and a prescription for identifying points."} Op.Cit
\cite {DD}.

The matching condition is just the screw translation in ${\Re}^3$,  one of the transitive actions of the automorphism of the Heisenberg group, modulo discrete sub-group ${\bf Z}$.

\section*{\bf Acknowledgements}I am grateful to S.Deser, S.Jofilly, L.Ryff, K.Mundim, A.Antunes, C.Sigaud and R.Araujo for discussions, and to Marcela Gon\c calves for continuing encouragements.

\end{document}